\documentclass{article}

\usepackage[utf8]{inputenc}
\usepackage[english]{babel}
\usepackage{latexsym}
\usepackage{graphicx}
\usepackage{hyperref}
\usepackage{amsmath}
\usepackage{authblk}
\usepackage{tocbibind}
\usepackage{geometry}
\usepackage{booktabs}
\usepackage{cite}
\usepackage{amsfonts}
\usepackage{hyperref}
\date{}
\title{The Evolution of Language in Social Media Comments}
\author[a]{Niccolò Di Marco}
\author[b]{Edoardo Loru}
\author[c]{Anita Bonetti}
\author[d]{Alessandra Olga Grazia Serra}
\author[a]{Matteo Cinelli}
\author[a,1]{Walter Quattrociocchi}

\affil[a]{Department of Computer Science, Sapienza University of Rome}
\affil[b]{Department of Computer, Control and Management Engineering, Sapienza University of Rome}
\affil[c]{Department of Communication and Social Research (CoRiS), Sapienza University of Rome}
\affil[d]{Tuscia University}

\begin{document}

\maketitle

\begin{abstract}
Understanding the impact of digital platforms on user behavior presents foundational challenges, including issues related to polarization, misinformation dynamics, and variation in news consumption. Comparative analyses across platforms and over different years can provide critical insights into these phenomena. This study investigates the linguistic characteristics of user comments over 34 years, focusing on their complexity and temporal shifts. Utilizing a dataset of approximately 300 million English comments from eight diverse platforms and topics, we examine the vocabulary size and linguistic richness of user communications and their evolution over time.
Our findings reveal consistent patterns of complexity across social media platforms and topics, characterized by a nearly universal reduction in text length, diminished lexical richness, but decreased repetitiveness. Despite these trends, users consistently introduce new words into their comments at a nearly constant rate.
This analysis underscores that platforms only partially influence the complexity of user comments. Instead, it reflects a broader, universal pattern of human behaviour, suggesting intrinsic linguistic tendencies of users when interacting online.
\end{abstract}

\noindent {\bf Keywords:} Social Media $|$ Language Evolution $|$ Social Dynamics

\maketitle

\section*{Introduction}

The rapid expansion of social media platforms has revolutionized how we connect and communicate, fundamentally altering the landscape of human interaction. These platforms have become integral to our daily lives as primary sources of information, entertainment, and personal communication \cite{tucker2018social,aichner2021twenty, DiMarco2024}. While they offer unprecedented opportunities for connectivity and interaction, they also intertwine entertainment-driven business models with complex social dynamics, raising substantial concerns about their impact on users and society at large \cite{guess2023social}. The influence of social media on public discourse and individual behavior has become a pressing concern within the scientific community, particularly regarding issues of polarization, misinformation \cite{Tucker2018,cinelli2021echo,gonzalez2023asymmetric,guess2023social, Falkenberg2022}, and hate speech \cite{castano2021internet,lupu2023offline,siegel2020online}. Recent studies have explored their potential effects on user behavior, uncovering complex interactions and identifying key unresolved questions \cite{Avalle2024,gonzalez2023asymmetric, guess2023social,nyhan2023like,guess2023reshares}. They confirm that online users often select information that aligns with their preferences, overlook dissenting information, and form homophilic communities \cite{cinelli2021echo}, which may influence their belief formation and communication methods. In this landscape, investigating the textual properties and the vocabulary of user-generated content is crucial for understanding evolving social dynamics in terms of lexicon influence \cite{greaves2013use,xu2019sentiment,alrumaih2020sentiment,volkova2015inferring}.

Historically, measuring vocabulary has posed a significant challenge in psychology, considering its close association with other cognitive skills such as reading comprehension and information processing \cite{smith1941measurement,tschirner2004breadth}. While previous studies have extensively examined the impact of vocabulary size on academic success, revealing substantial individual differences and emphasizing the pivotal role of lexical knowledge in educational settings \cite{milton2013vocabulary}, there is still a gap in understanding how these dynamics adapt to the digital era. Indeed, linguistics is increasingly focusing on social media, spurred by concerns that internet language may corrupt traditional writing practices and face-to-face communication \cite{baron2008always}. Language adapts to changes in culture, society, and technology, leading to the emergence of novel linguistic forms like abbreviations, phonetic spellings, neologisms, and multimedia elements such as hashtags and emojis \cite{eisenstein2014diffusion}.

However, a systematic understanding of language complexity, despite extensive discourse, still needs to be improved.
Current research often addresses specific elements of complexity but needs to provide a holistic view, making it challenging to develop a unified theoretical framework. The definition of complexity within linguistic studies is often ambiguous \cite{gong2011report}, and the role of quantitative metrics in assessing complexity has not been fully explored, with only limited studies addressing this approach \cite{ehret2023measuring,mirzapour2020measuring}. Scholarly debate on linguistic complexity tends to adopt two distinct perspectives. The first views complexity as a theoretical abstraction with little direct application to real-world linguistic scenarios. The second treats it as an empirical phenomenon that can be quantified and analyzed using theoretical tools. Miestamo introduces the concepts of `absolute complexity' and `relative complexity' \cite{miestamo2008grammatical}. Absolute complexity is considered an intrinsic property of language systems, independent of user interaction. In contrast, relative complexity involves the user's perspective, measuring complexity as the cost or difficulty encountered by language users. 

Our research primarily explores relative complexity, focusing on how it manifests in user interactions on social media platforms. In more detail, this study investigates whether and how social media platforms have influenced the linguistic constructs of user communications online. We analyze a large dataset of nearly 300 million English comments across eight major social media platforms—Facebook, Twitter, YouTube, Voat, Reddit, Usenet, Gab, and Telegram—covering nearly three decades and several topics.
We begin by assessing the average vocabulary size of users according to their activity level. We then provide a framework to evaluate the speed at which users reach their maximum vocabulary, offering new insights into how digital environments shape communication norms and user engagement. Finally, we map text complexity by applying two established lexical richness and repetitiveness metrics and we explore their evolution over several years.

\section*{Users vocabulary in social media}
In the following sections, we conduct a comparative analysis of 8 different social media platforms: Facebook, Twitter, YouTube, Voat, Reddit, Usenet, Gab and Telegram. Each social media contains comments related to different topics (see Data Collection section for further details).

\subsection*{The vocabulary of users}
To measure the vocabulary size on social media platforms, we aggregate all the comments from each user into one document. We then perform text preprocessing, including tokenization and stemming, to facilitate accurate word counting (see further details about this procedure in Materials and Methods section). In this work, we refer to {\it tokens} as instances of individual words as they appear in the text, whereas {\it types} represent distinct words without repetition. Therefore, we associate to each user a couple of integer values containing the number of tokens (i.e., the number of their total words) and the number of types (i.e., the number of their unique words).

\begin{figure}
    \centering
    \includegraphics[width = .9\linewidth]{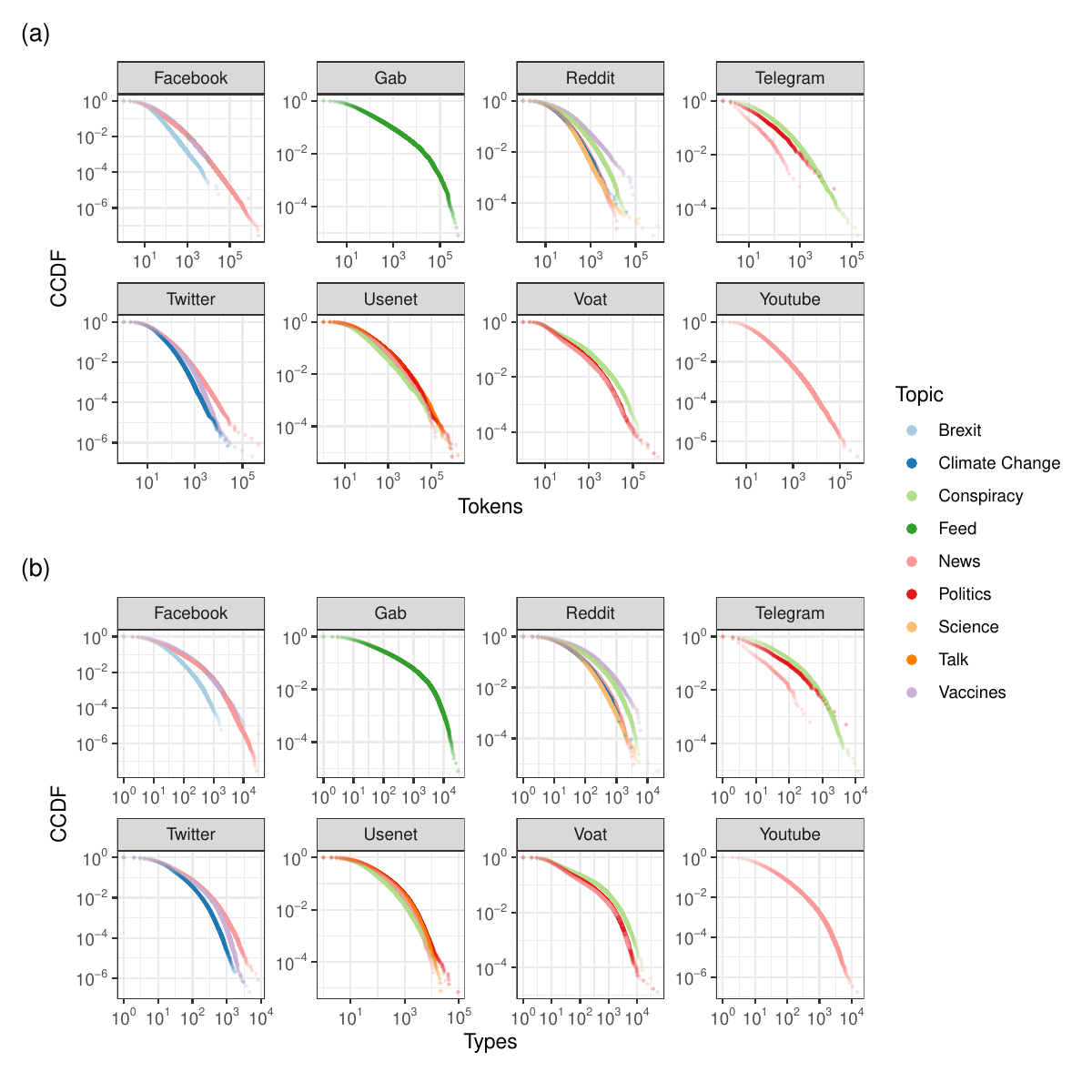}
    \caption{CCDF of the distributions of number of $(a)$ tokens and $(b)$ types used by each user.}
    \label{fig:types_tokens_dist}
\end{figure}

Figure \ref{fig:types_tokens_dist} displays the complementary cumulative distribution functions (CCDF) of the number of tokens and types within the documents of each user in each dataset. 

The distributions display general consistency across different social media platforms and topics, albeit with varying magnitudes. Their behaviors are almost identical, with the primary differences manifesting in their tails, since tokens exhibit longer tails. This pattern aligns with the expectation that shorter texts typically have nearly identical numbers of words and unique words, whereas longer ones can present richer vocabularies.
This observation is further supported by Figure \ref{fig:TTR_dist} in Supplementary Information (SI), which illustrates the distributions of Type-Token Ratios (TTR). These distributions are predominantly peaked at 1, indicating a high similarity between types and tokens in most cases, with only a minority of users exhibiting lower TTR values.
Nonetheless, consistently across various topics and social media, most users typically employ up to 10 unique words, indicating a relatively small vocabulary size. 

\begin{figure}
    \centering
    \includegraphics[width = .8\linewidth]{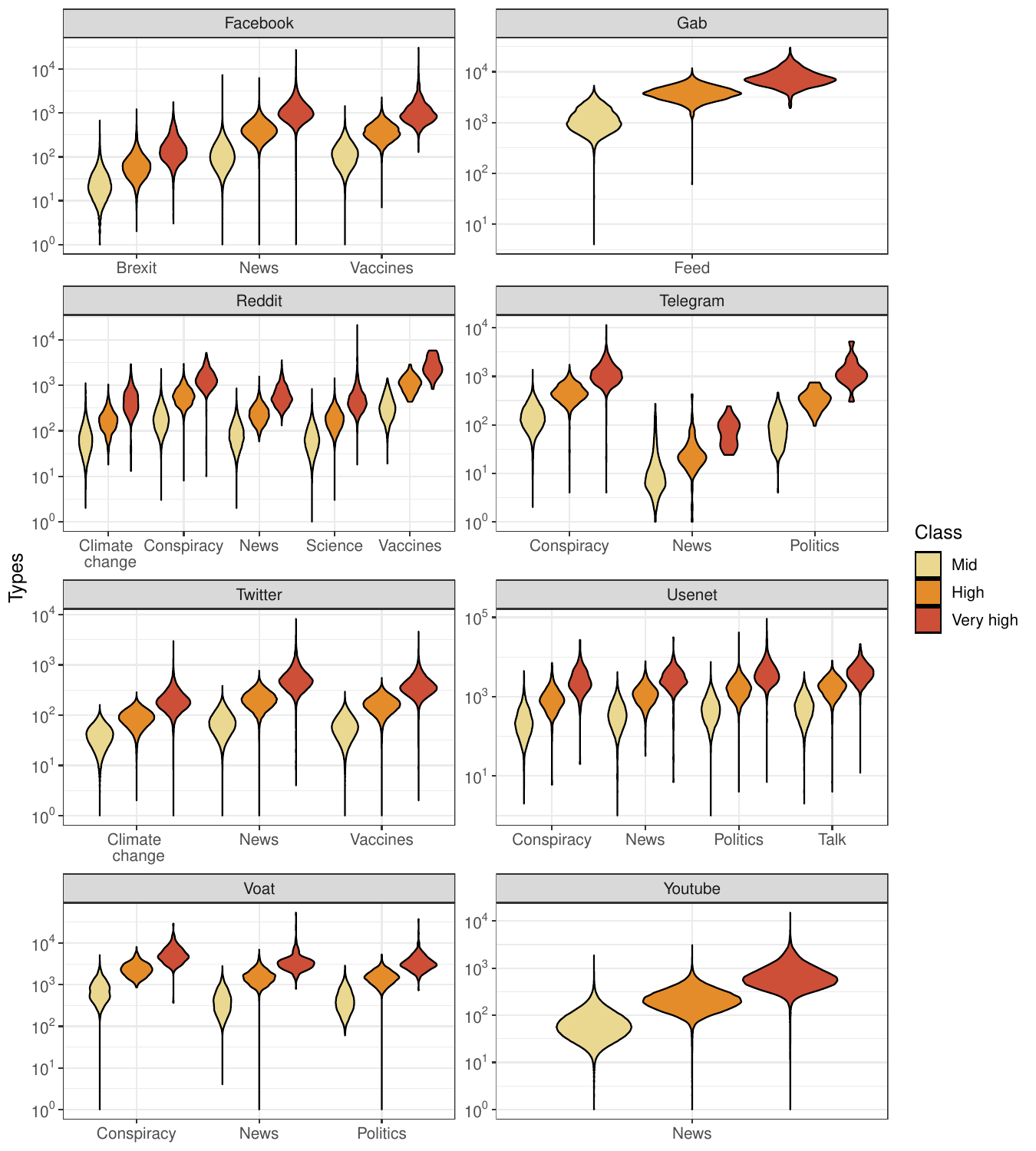}
    \caption{Distribution of the number of types (i.e., unique words) employed by users according to their activity class, which is determined by the number of comments they left in each specific dataset.}
    \label{fig:types_by_class}
\end{figure}

Obviously, this observation is somewhat dependent on the user activity, which is known to follow a heavy-tailed distribution \cite{Avalle2024} (i.e., only a few users exhibit high activity, while the majority show very low participation), potentially skewing the observed vocabulary sizes.
To disentangle the possible effect of user activity, we categorize users into four classes —{\it low, medium, high}, and {\it very high}—based on the number of comments they have posted on each specific dataset (more details about the classification criteria are provided in the Materials and Methods section).
Figure \ref{fig:types_by_class} illustrates the distribution of {\it types} within each class except {\it Low}, thus considering only users with sufficient activity.

\begin{figure*}[!ht]
    \centering
    \includegraphics[width = .9\linewidth]{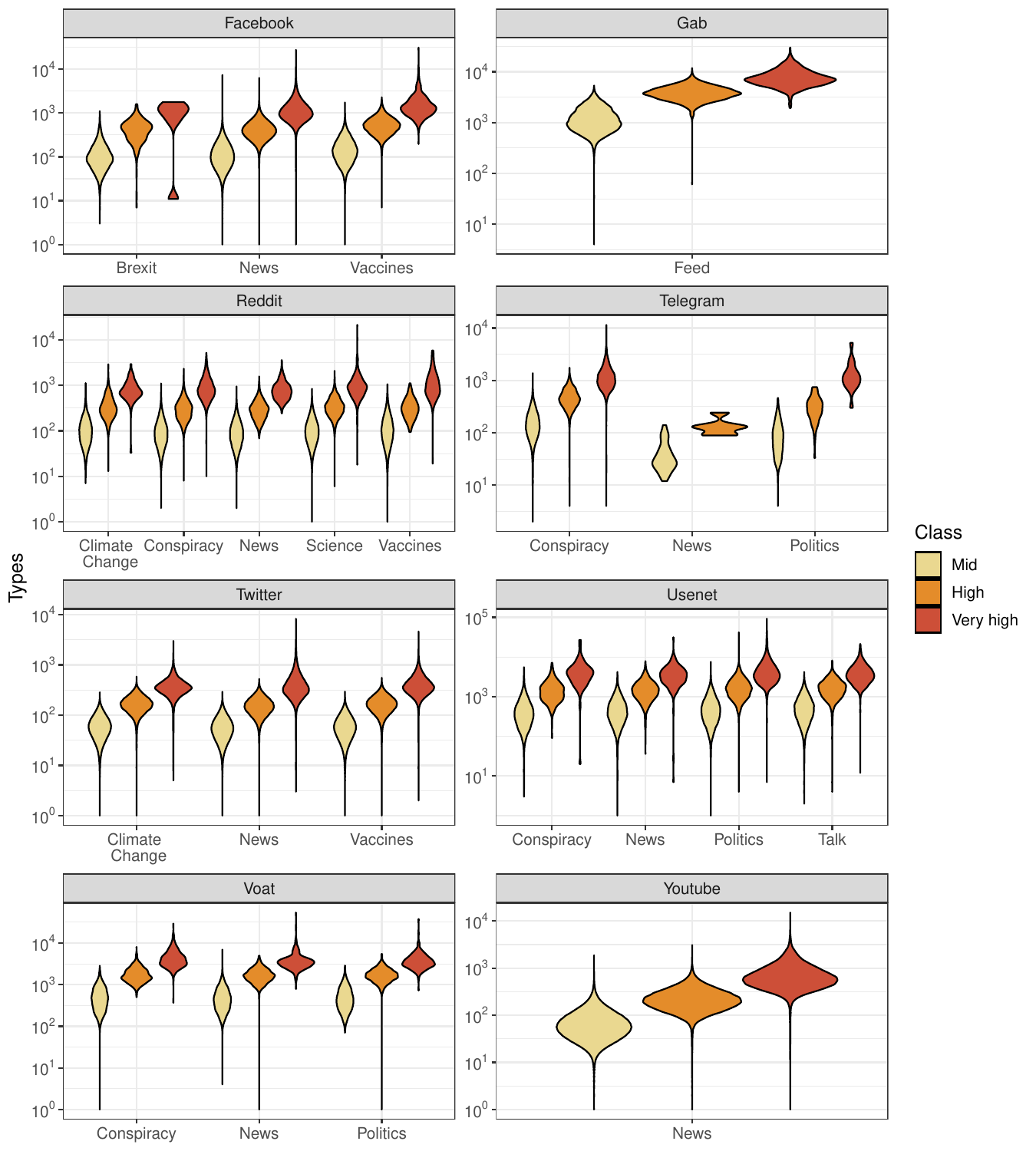}
    \caption{Distribution of the number of types (i.e., unique words) employed by users according to their activity class. In this case, the activity classes are computed using the comments distribution of the whole social platform.}
    \label{fig:types_class}
\end{figure*}

As expected, higher activity levels among users are associated with vocabulary distributions centered around higher values, indicating a more diverse lexical usage.
Interestingly, while vocabulary size distributions generally show consistency across various social media platforms, it seems specific topics demand larger vocabularies. However, the shifting in the distributions could be a consequence of the different sizes of each dataset, resulting in some users having a larger number of comments and thus a larger number of types/tokens. To control for this effect, Figure \ref{fig:types_class} in SI represents the same plots but in which activity classes are computed according to the whole distribution of comments of a specific social. Thus, we are able to compare the behaviour of users with approximately the same number of comments.
The results show that the shifting disappear, providing evidence that neither the topic of discussion can influence the total number of types used by users.

Furthermore, we examine whether users' vocabulary distributions adhere to Zipf's Law \cite{zipf2013psycho,zipf2016human}, a principle that suggests a predictable frequency distribution of words in human languages. The results, in SI, indicate consistent exponents across social media and topics, supporting the hypothesis that the form of Zipf's Law is a valid approximation of the frequency term observed in online social media communications.

\subsection*{Vocabulary Evolution}
In the previous section we have explored the aggregated users' production without examining their evolution. To address this gap, we analyze how the vocabularies of individual users evolve over time by tracking the rate at which they introduce new words in their time-ordered comments.

We chronologically arrange each user's comments and apply the same tokenization process used for our previous analyses. 

For a user $u$ with $n$ comments, we compute the vector $\mathbf{v}^u \in \mathbb{N}^n$, where each entry $i$ represents the cumulative count of unique words up to the $i$-th comment. 

For example, if a user $u$ writes two comments, the first containing the words ``politics," ``health," ``comics," and the second ``politics," ``left," ``right," then $\mathbf{v}^u$ would be (3, 5). We calculate $\mathbf{v}^u$ for all users having 25 to 100 comments (for manageability we consider 50 to 100 comments in Facebook News). This range ensures that we focus on users with a reasonable level of activity and excludes accounts that may be malicious or not genuine, often exhibiting excessively high comment counts.

We employ a linear interpolation on the scaled values of $\mathbf{v}$ in $[0,1]$ to track the evolution of users' vocabularies. Since $v_i \leq v_{i+1}, i = 1, \ldots, N$, a possible measure of the speed at which each user reaches its maximum vocabulary is the area under this curve, similar to methodologies used in previous studies \cite{fb_engagement}. 
This value, which ranges from 0 to 1, provides insights into vocabulary usage patterns: values close to 0 indicate a late expansion of vocabulary, values close to 1 suggest rapid saturation of vocabulary usage early on, and values around 0.5 indicate a steady increase in vocabulary across comments.

\begin{figure}[!ht]
    \centering
    \includegraphics[width = \linewidth]{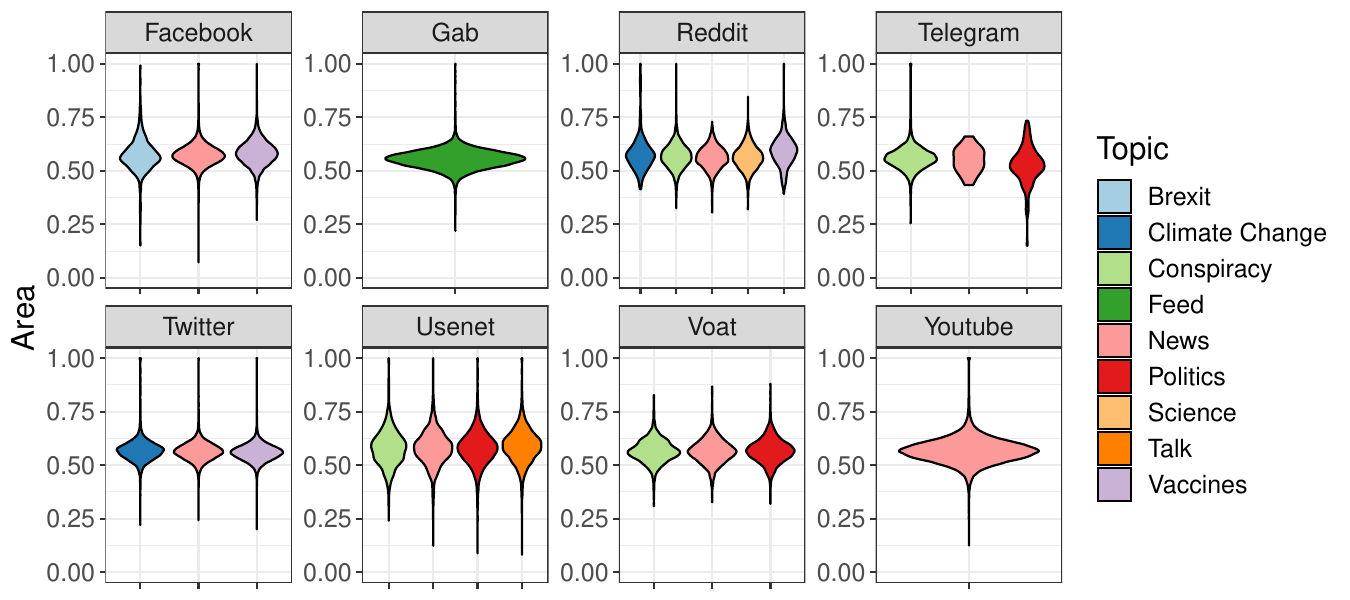}
    \caption{Distribution of the area under the curves determined by users' progressive exploration of their vocabulary.}
    \label{fig:area}
\end{figure}

Figure \ref{fig:area} illustrates the distribution of these measurements across different topics and social media platforms.

We observe a general consistency across all distributions, peaking at values around $0.6$; this suggests, on average, a continuous but modest addition of new words to a user's vocabulary with a minority of users reaching their whole vocabulary in the first comments.

The consistency of the findings underscores universal behaviors that appear to be largely independent of the specific platforms and topics involved.

\subsection*{Comments complexity}
Beyond the size of users' vocabularies, exploring the general complexity of comments adds a significant dimension to our analysis.
The complexity of texts can be approached from various perspectives, and the literature made numerous metrics available \cite{golcher2007new,yule2014statistical,Tweedie1998,dugast1979vocabulaire,zipf2013psycho,renyi1961measures,TanakaIshii2015}. After a careful review of the available measures and previous works, we decided to rely upon two of them that are able to provide a rather orthogonal perspective in terms of text complexity, namely Yule's $K$-complexity and gzip complexity $g$. The detailed methodologies are described in the Materials and Methods section.
We recall that a high $K$ suggests a small lexical complexity, whereas it assumes its minimum (i.e., $K = 0$) for a text in which only distinct words are used. On the other hand, values of $g$ close to $0$ (or even negative) suggest texts with low repetitiveness, while values close to $1$ indicate texts with high repetitive patterns.  

\begin{figure}[!ht]
    \centering
    \includegraphics[width = \linewidth]{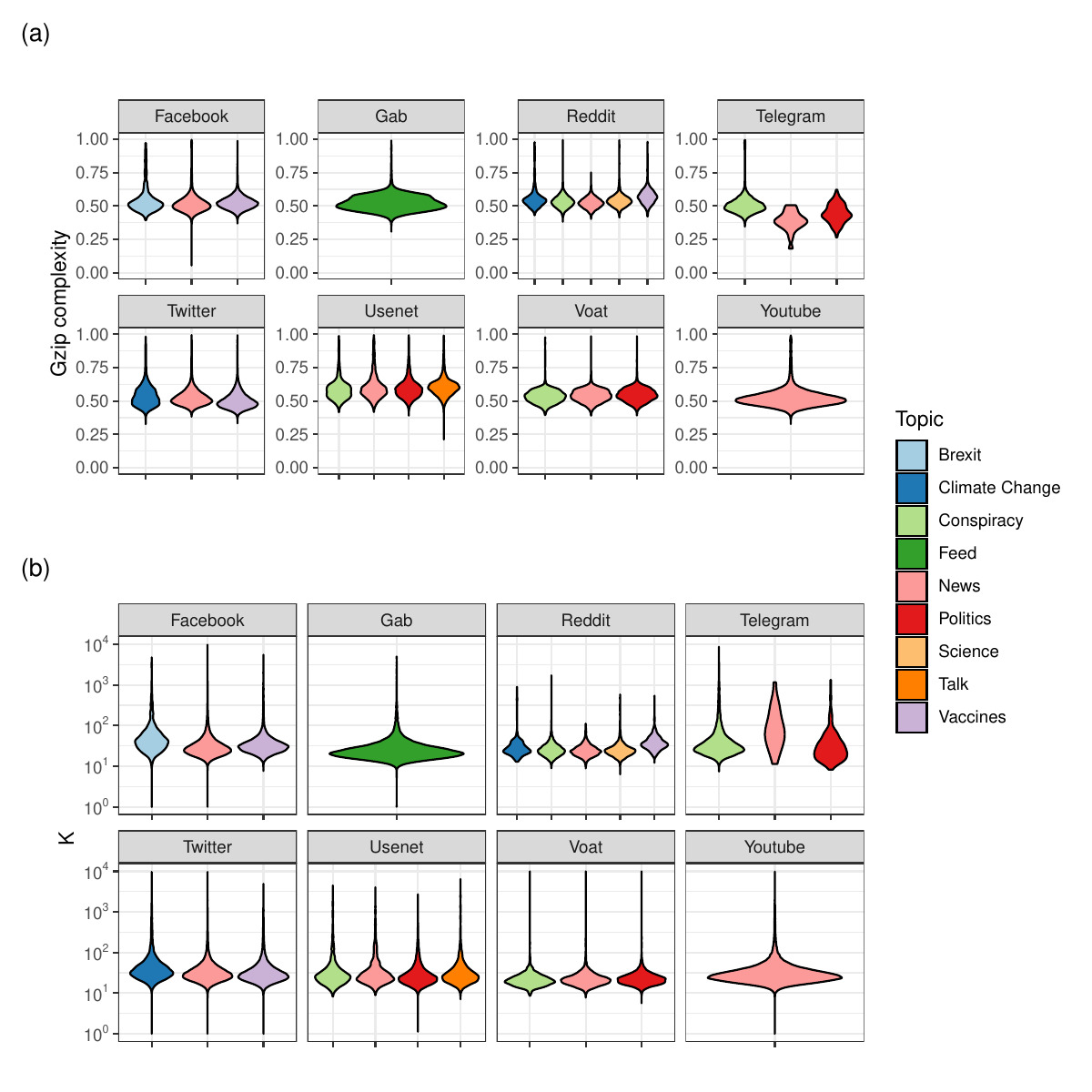}
    \caption{Distribution of $(a)${\it gzip} complexity and $(b)$ K-complexity for users having at least 20 comments. 
    For the larger dataset, we selected a sample of 50000 users to compute $K-$complexity. For visual reasons, we add 1 to all the values of $K-$complexity.}
    \label{fig:complexity_dist}
\end{figure}

As with our previous analysis, we compile the complete set of comments from each user into single documents. $K$-complexity is calculated following the previously outlined preprocessing steps, whereas gzip complexity is assessed using the raw texts, as done by previous works \cite{Lee_Eoff_Caverlee_2021}.
Figure \ref{fig:complexity_dist} illustrates the distribution of both complexity measures among users who have posted at least 20 comments. For datasets comprising over 50,000 users, Yule's $K$-complexity was calculated on a sample of 50,000 users to ensure manageability and computational efficiency.

We observe consistency in the distributions of both $K$ and $g$ on different social media platforms and topics. Generally, the texts produced by users display moderate lexical complexity and repetitiveness. However, a minority of users produce highly repetitive texts and exhibit low complexity, which may indicate the presence of automated or coordinated accounts \cite{di2024post,cinelli2022coordinated, nwala2023, keller2019, pacheco2021, Lee_Eoff_Caverlee_2021}. 

In SI, we present an analysis where each user's comments are randomized before being aggregated into documents, i.e. the document associated with a user is made of random comments from other users. The results show distributions with a lower variance, suggesting much more uniform behaviors. Although distributions look similar, Mann-Whitney tests detect that the real and null distributions are different in almost all cases, suggesting that users adopt their vocabulary, that is not replicable with a random assignment of comments.
    
\subsection*{Evolution of complexity} 
In previous sections, we have explored the complexity of texts written by users online. However, an interesting point to investigate regards the evolution of the complexity of comments over time as a proxy for the use of social media as a public square for opinion sharing and active debate. Thus, our focus now shifts to determining whether the complexity of comments has changed over time.
To achieve this, we select subsets of datasets with a sufficiently broad time span. Specifically, we include Facebook News, Facebook Vaccines, Twitter Vaccines, Usenet News, Usenet Politics, Usenet Talk, Voat News, Voat Politics, and YouTube News and we analyze the whole set of comments, without aggregating by user.

\begin{figure}[!ht]
    \centering
    \includegraphics[width = \linewidth]{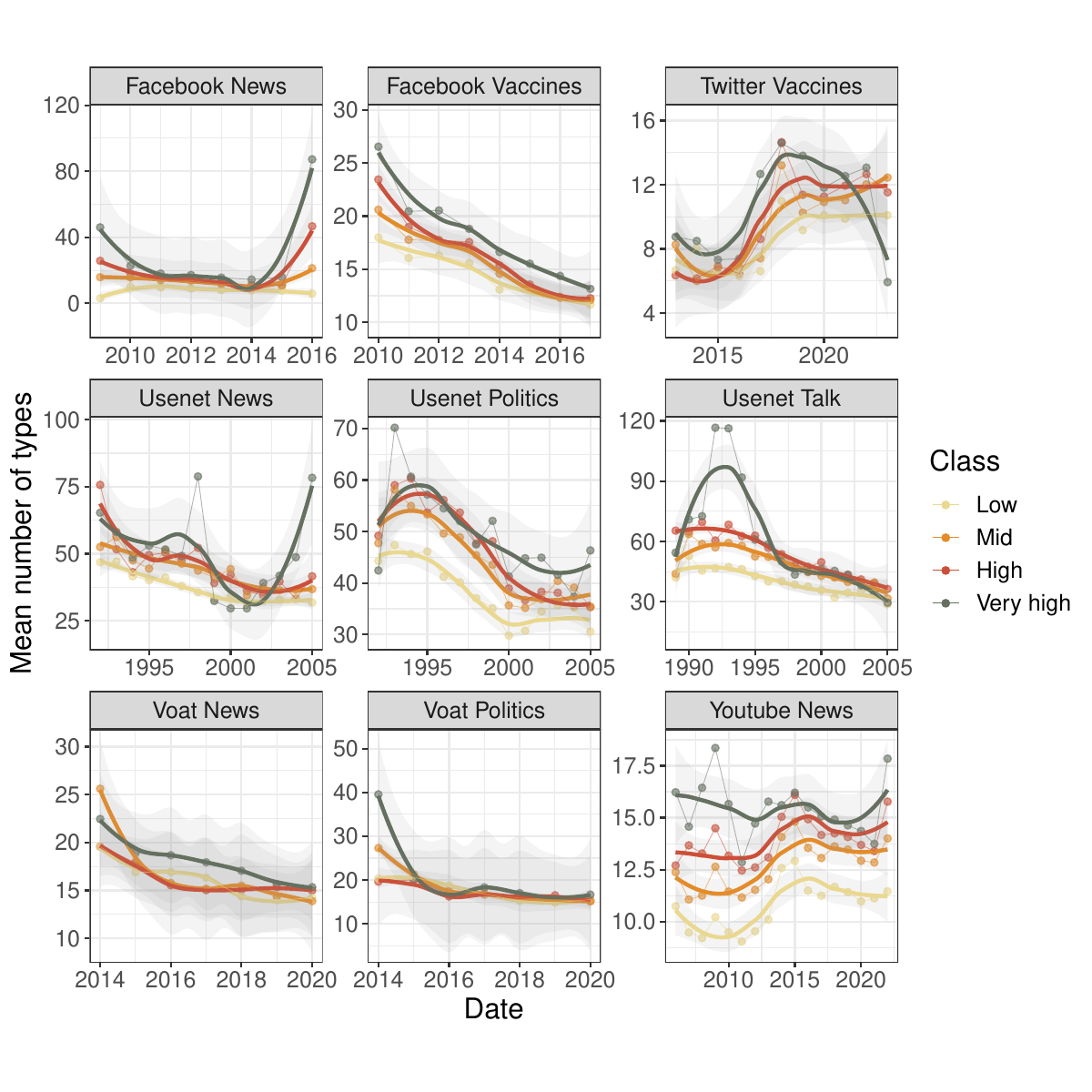}
    \caption{Evolution of the mean number of types in each dataset. The smooth curves are obtained using a {\it loess} regression}
    \label{fig:evolution_types}
\end{figure}

First, we examine the evolution of the number of types (i.e., unique words), being the simplest complexity measure. To mitigate any potential bias from users' activity levels, for each year we classify each user into one of four activity classes —low, mid, high, or very high— as previously established.
Figure \ref{fig:evolution_types} illustrates the annual progression of the average number of types across all activity classes, computed for all years with a minimum number of 100 comments.

\begin{figure*}[!ht]
    \centering
    \includegraphics[width = .8\linewidth]{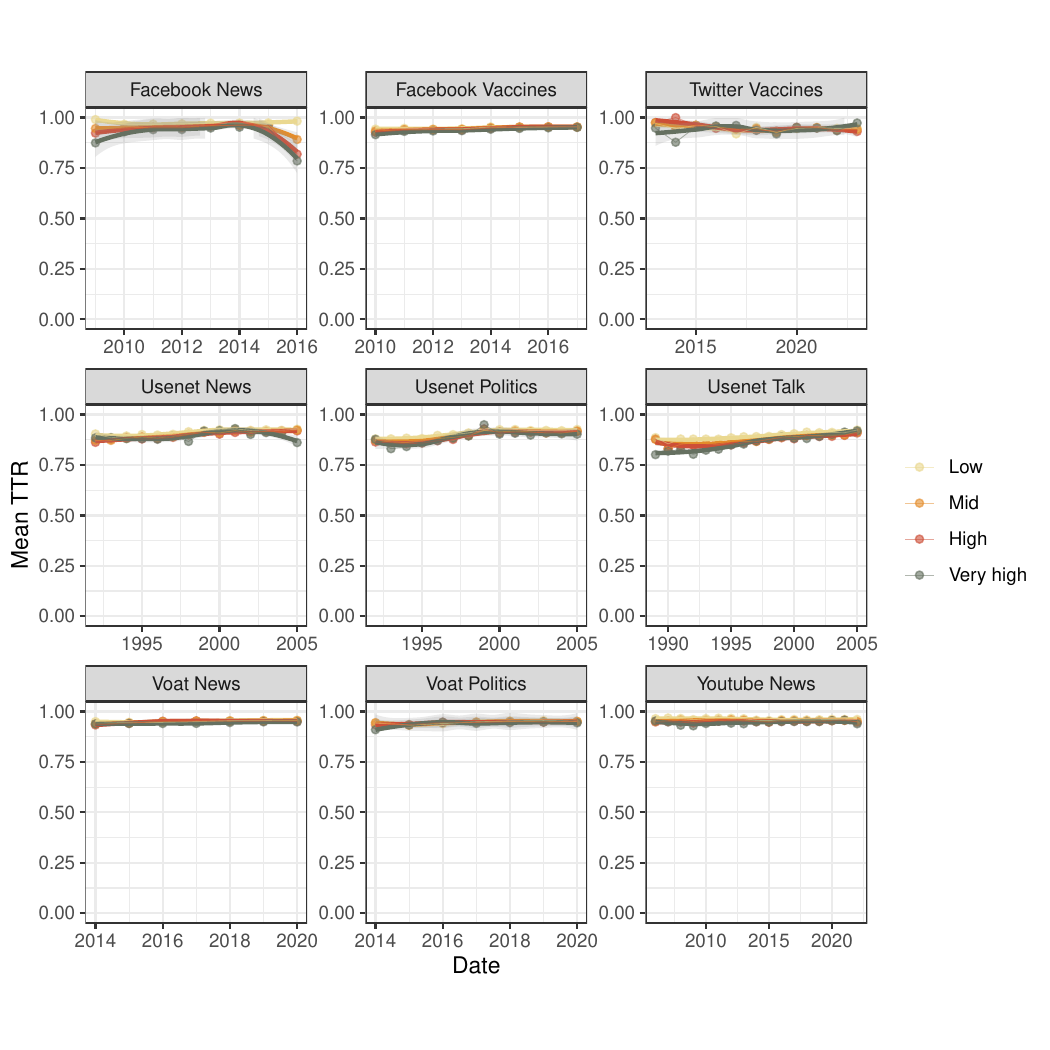}
    \caption{Evolution of TTR values across datasets and classes of activities.}
    \label{fig:TTR_time}
\end{figure*}

The number of unique words appears to decrease in all user classes except for those on Twitter and YouTube. Additionally, Figure \ref{fig:TTR_time} in SI reveals that the Type-Token Ratio (TTR) remains relatively stable across all platforms, suggesting a concurrent reduction in the total number of words (i.e., tokens) used. This trend reflects also a decrease in user activity, characterized by shorter comments that contain fewer unique words.

To further quantify the relationship between time and text complexity, we implement a regression model with interaction terms to account for the specific social media platform where each comment was posted. For this analysis, we use a sample of $6000$ comments per year from each platform and topic, ignoring all years having less than this number of comments.

We provide a detailed discussion of the model in the Materials and Methods section. Moreover, Table S1 contains a breakdown of the dataset used for this experiment.

We recall here that our regression model employs complexity measures as regressors and the year of the comments as the dependent variable. To allow for a better comparison between measures, we first normalize each regressor and, after detecting a heteroskedasticity problem, we correct standard errors with their robust version. The model, whose coefficients are detailed in Table \ref{tab:regression}, achieves an adjusted $R^2$ value of approximately 0.87.

\begin{table*}[!ht]
\centering
\begin{tabular}{lrrrr}
  \toprule
Variables & Estimate & Total estimate & Standard error & $p$ \\ 
  \midrule
$\beta_0$ & 2013.091 & 2013.091 & 0.010 & $<0.001$ \\ 
  $\beta_1$ & -0.367 & -0.367 & 0.034 & $<0.001$ \\ 
  $\beta_2$ & 0.079 & 0.079 & 0.007 & $<0.001$ \\ 
  $\beta_3$ & -0.228 & -0.228 & 0.007 & $<0.001$ \\ 
  $\beta_{0,tw}$ & 7.344 & 2020.435 & 0.036 & $<0.001$ \\ 
  $\beta_{0,un}$ & -13.644 & 1999.447 & 0.013 & $<0.001$ \\ 
  $\beta_{0,vt}$ & 4.376 & 2017.467 & 0.013 & $<0.001$ \\ 
  $\beta_{0,yt}$ & 2.076 & 2015.167 & 0.021 & $<0.001$ \\ 
  $\beta_{1,tw}$ & 4.548 & 4.181 & 0.136 & $<0.001$ \\ 
  $\beta_{1,un}$ & 0.314 & -0.054 & 0.035 & $<0.001$ \\ 
  $\beta_{1,vt}$ & 0.357 & -0.011 & 0.039 & $<0.001$ \\ 
  $\beta_{1,yt}$ & 1.031 & 0.664 & 0.064 & $<0.001$ \\ 
  $\beta_{2,tw}$ & -0.071 & 0.008 & 0.009 & $<0.001$ \\ 
  $\beta_{2,un}$ & -0.067 & 0.012 & 0.012 & $<0.001$ \\ 
  $\beta_{2,vt}$ & -0.067 & 0.012 & 0.010 & $<0.001$ \\ 
  $\beta_{2,yt}$ & -0.070 & 0.009 & 0.019 & $<0.001$ \\ 
  $\beta_{3,tw}$ & -0.098 & -0.326 & 0.032 & $0.002$ \\ 
  $\beta_{3,un}$ & -0.540 & -0.769 & 0.015 & $<0.001$ \\ 
  $\beta_{0,vt}$ & 0.047 & -0.181 & 0.012 & $<0.001$ \\ 
  $\beta_{0,yt}$ & 0.169 & -0.059 & 0.021 & $<0.001$ \\ 
   \bottomrule
\end{tabular}
\caption{Results of regression model.}
\label{tab:regression}
\end{table*}

We use Facebook as the baseline, i.e. the estimates of coefficients $\beta_i$ refer to Facebook comments. Moreover, the `Total Estimate' column reports the sum of the coefficient with the relative baseline, i.e. if $\beta_{k,j}$ is considered, the column reports the value $\beta_{k,j} + \beta_k$, thus describing the change in $y$ after a unit change of regressor $k$ if all the other variables are fixed and the comment comes from social $j$. In other words, $\beta_{k,j}$ is the deviation from the behavior observed on Facebook.

The results suggest that the number of unique words is positively correlated with more recent years on Twitter and YouTube, as previously observed. Conversely, there is a negative relation on other platforms.
Regarding Yule's K-complexity, higher values are associated with more recent years, thus suggesting that comments exhibit a decrease in lexical complexity over time. 
Finally, $g$-complexity consistently shows negative associations with time across all platforms, indicating that recent comments tend to have lower repetitiveness, despite their lower lexical complexity.

To check the robustness of the results, we conduct a similar analysis in SI using a logistic regression. Notably, despite the use of a different methodology, the results show consistency.

Overall, the data indicates that comments across platforms exhibit lower lexical complexity and repetitiveness as time progresses. Furthermore, in nearly all cases, comments become shorter and contain fewer unique words.

\section*{Conclusions}
Our comprehensive analysis across eight major social media platforms reveals consistent patterns in user behavior and language complexity. This study clarifies how users' linguistic behavior has adapted to the digital era and evolved over nearly three decades of internet use. Despite the diversity of topics and platforms analyzed, which encompass approximately 300 million English comments over nearly three decades, we find a general decrease in the length of comments and a reduction in lexical richness. Notably, our findings suggest that language complexity is not influenced by the platform, but rather reflect a broader, universal aspect of human communication. We have also explored how individual users evolve their vocabulary over time, discovering that most users gradually introduce new words, with distributions peaking at relatively low levels of vocabulary expansion. This study underscores the persistent nature of linguistic behavior across digital communication platforms, emphasizing the influence of inherent human communication traits over technological or topical variations. Our results provide valuable insights into the dynamics of language use in digital environments, supporting the hypothesis of universal linguistic patterns among social media users.

\section*{Materials and Methods}
\subsection*{Data collection}
Below we report the detailed procedure to gather each dataset.

\textbf{Facebook} - We utilize datasets from prior studies on discussions concerning Vaccines \cite{MMschmidt2018polarization}, News \cite{MMschmidt2017anatomy}, and Brexit \cite{MMdelvicario2017brexit}. For the vaccine topic, the dataset comprises approximately 2 million comments from public groups and pages collected over the period from January 2, 2010, to July 17, 2017. For the News topic, we selected pages from the Europe Media Monitor that reported the news in English, resulting in a dataset containing roughly 362 million comments between September 9, 2009, and August 18, 2016. Additionally, this dataset includes approximately 4.5 billion likes associated with posts and comments on these pages. Lastly, for the Brexit topic, the dataset encompasses around 460,000 comments from December 31, 2015, to July 29, 2016.

\textbf{Gab} - We collected data from the Pushshift.io archive (\url{https://files.pushshift.io/gab/}) on discussions from the platform's inception on August 10, 2016, until it temporarily went offline on October 29, 2018, following the Pittsburgh shooting \cite{reuters2018gab}. The dataset includes approximately 14 million comments.

\textbf{Reddit} - Data was collected from the Pushshift.io archive (\url{https://pushshift.io/}) covering the period from January 1, 2018, to December 31, 2022. We manually identified and selected subreddits for each topic that best represented the targeted discussions. From this process, we gathered approximately 800,000 comments from the \textit{r/conspiracy} subreddit for the Conspiracy topic. For the Vaccines topic, we collected about 70,000 comments from the \textit{r/DebateVaccines} subreddit, focusing on the COVID-19 vaccine debate. The \textit{r/News} subreddit provided roughly 400,000 comments for the News topic. From the \textit{r/environment} subreddit, we obtained approximately 70,000 comments related to the Climate Change topic. Lastly, the \textit{r/science} subreddit yielded about 550,000 comments for the Science topic.

\textbf{Telegram} - We compiled a list of 14 channels, each linked to one of the study's topics. We manually collected messages and their associated comments from each channel. From the 4 channels related to the News topic (\textit{news\_notiziae, news\_ultimora, news\_edizionestraordinaria, news\_covidultimora}), we gathered approximately 724,000 comments from posts dated between April 9, 2018, and December 20, 2022. For the Politics topic, the 2 channels (\textit{politics\_besttimeline, politics\_polmemes}) yielded a total of about 490,000 comments from the period between August 4, 2017, and December 19, 2022. Lastly, the 8 channels focused on the Conspiracy topic (\textit{conspiracy\_bennyjhonson, conspiracy\_tommyrobinsonnews, conspiracy\_britainsfirst, conspiracy\_loomeredofficial, conspiracy\_thetrumpistgroup, conspiracy\_trumpjr, conspiracy\_pauljwatson, conspiracy\_iononmivaccino}) produced approximately 1.4 million comments from August 30, 2019, to December 20, 2022. 

\textbf{Twitter} - We utilized datasets from previous research that include discussions on Vaccines \cite{MMvalensise2021lack}, Climate Change \cite{Falkenberg2022}, and News \cite{MMquattrociocchi2022reliability}. We collected approximately 50 million comments on the Vaccines topic from January 23, 2010, to January 25, 2023. For the News topic, we expanded the dataset used in \cite{MMquattrociocchi2022reliability} by including all threads with fewer than 20 comments, resulting in a total of approximately 9.5 million comments collected from January 1, 2020, to November 29, 2022. Lastly, we gathered about 9.7 million comments on the Climate Change topic from January 1, 2020, to January 10, 2023. 

\textbf{Usenet} - We collected data from the Usenet discussion system using the Usenet Archive (\url{https://archive.org/details/usenet?tab=about}). We identified a range of topics, including extensive, broad, and heterogeneous discussions within active and populated newsgroups. As a result, we selected conspiracy, politics, news, and talk as the topic candidates for our analysis. We gathered approximately 280,000 comments from the \textit{alt for the Conspiracy topic.conspiracy} newsgroup, from September 1, 1994, to December 30, 2005. About 2.6 million comments were collected from the \textit{alt for the Politics topic.politics} newsgroup between June 29, 1992, and December 31, 2005. We obtained approximately 620,000 comments for the News topic from the \textit{alt.news} newsgroup, from December 5, 1992, to December 31, 2005. Finally, we collected all discussions from the \textit{alt for the Talk topic.talk} newsgroup, totaling about 2.1 million contents, from February 13, 1989, to December 31, 2005.

\textbf{Voat} - We utilized a dataset described in \cite{MMmekacher_voat_2022} that encompasses the entire lifespan of the platform from January 9, 2018, to December 25, 2020. This dataset includes approximately 16.2 million posts and comments from around 113,000 users across approximately 7,100 subverses (Voat's equivalent of a subreddit). Similarly to previous platforms, we linked topics to specific subverses. As a result, for the Conspiracy topic, we collected about 1 million comments from the \textit{greatawakening} subverse during the platform's operational period. For the Politics topic, we gathered roughly 1 million comments from the \textit{politics} subverse between June 16, 2014, and December 25, 2020. Lastly, we amassed approximately 1.4 million comments from the \textit{news} subverse for the News topic between November 21, 2013, and December 25, 2020.

\textbf{YouTube} - We utilized a dataset referenced in previous research that initially focused on Climate Change discussions \cite{Falkenberg2022}. This dataset has been expanded to include conversations on Vaccines and News topics, following the approach used for other platforms. The data collection for YouTube was conducted using the YouTube Data API (\url{https://developers.google.com/youtube/v3}). For the Climate Change topic, we collected approximately 840,000 comments from March 16, 2014, to February 28, 2022. For the Vaccines topic, we gathered comments between January 31, 2020, and October 24, 2021, that include keywords related to COVID-19 vaccines, such as \emph{Sinopharm, CanSino, Janssen, Johnson\&Johnson, Novavax, CureVac, Pfizer, BioNTech, AstraZeneca, Moderna}, resulting in about 2.6 million comments. Finally, for the News topic, we collected roughly 20 million comments from February 13, 2006, to February 8, 2022, including videos and comments from a list of UK-based news outlets, provided by Newsguard, a fact-checking agency.

We select only English text from this data as detected by the {\it cld3} package of R \cite{cld3}.
Table \ref{tab:data_breakdown} shows a breakdown of the resulting dataset. 

\begin{table}[!ht]
\centering
\begin{tabular}{llrr}
  \toprule
Dataset & Time range & Comments & Users \\ 
  \midrule
Facebook Brexit & 2015-12-31 2016-07-29 & 322365 & 171054 \\ 
  Facebook Vaccines & 2010-01-02 2017-07-17 & 1590907 & 304706 \\ 
  Facebook News & 2009-09-09 2016-08-17 & 229915622 & 36096691 \\ 
  Gab Feed & 2016-08-10 2018-10-29 & 10799968 & 126351 \\ 
  Reddit Climate change & 2018-01-01 2022-12-12 & 60113 & 22822 \\ 
  Reddit Conspiracy & 2018-01-01 2022-10-31 & 649054 & 82733 \\ 
  Reddit News & 2018-01-01 2018-12-31 & 358594 & 102982 \\ 
  Reddit Science & 2018-01-01 2022-12-11 & 488963 & 192675 \\ 
  Reddit Vaccines & 2018-01-21 2022-11-06 & 59980 & 4866 \\ 
  Telegram Conspiracy & 2019-08-30 2022-12-20 & 1111479 & 107009 \\ 
  Telegram News & 2018-05-24 2022-12-16 & 4493 & 1674 \\ 
  Telegram Politics & 2017-08-05 2022-12-19 & 20851 & 2013 \\ 
  Twitter Climate Change & 2020-01-01 2023-01-10 & 3562447 & 1467783 \\ 
  Twitter News & 2020-01-01 2022-11-29 & 5882655 & 1237679 \\ 
  Twitter Vaccines & 2010-01-23 2023-01-25 & 17682887 & 4916617 \\ 
  Usenet Conspiracy & 1994-09-01 2005-12-30 & 160632 & 30223 \\ 
  Usenet News & 1992-12-05 2005-12-30 & 385404 & 51204 \\ 
  Usenet Politics & 1992-06-29 2005-12-30 & 1541803 & 142469 \\ 
  Usenet Talk & 1989-02-13 2005-12-30 & 1390574 & 128321 \\ 
  Voat Conspiracy & 2018-01-09 2020-12-25 & 828018 & 24666 \\ 
  Voat News & 2013-11-21 2020-12-25 & 1164549 & 78199 \\ 
  Voat Politics & 2014-06-19 2020-12-25 & 902419 & 59442 \\ 
  Youtube News & 2006-02-13 2022-02-10 & 20946472 & 5623730 \\ 
   \bottomrule
\end{tabular}
\caption{Data breakdown of the dataset. We consider approximately 300M comments wrote by 50M users.}
\label{tab:data_breakdown}
\end{table}

\subsection*{Preprocessing of comments}
Before the main analysis, we tokenize the comments removing punctuation, symbols (and emojis), numbers, URL, hashtags, and English stopwords. Subsequently, we apply a stemmer to the resulting tokens, reducing them to their root form. 
After these steps, a text may remain empty (for example if it contains only tags or hashtags), therefore we remove all the comments containing 0 tokens.

The preprocessing and the main analysis were conducted using the {\it quanteda} package of R \cite{quanteda,quanteda.textstats}.

\subsection*{Measures of text complexity}
In linguistics, special attention has always been posed to develop measures capable of detecting the complexity of text. In particular, a text can be considered complex from a variety of points of view, such as lexical, readability, or repetitiveness. Many measures have been proposed (see \cite{quanteda.textstats} for a collection of measures provided by {\it quanteda.textstats} package in R \cite{quanteda}). Here, we have focused on lexical complexity and repetitiveness measures, using Yule's $K$ and $gzip$ complexity.

For what concerns the former, previous works have highlighted that many lexical complexity measures are incapable of being independent of text length \cite{TanakaIshii2015,Tweedie1998}. The same studies also highlight that, even if with some limitations, the well-known Yule's K-complexity \cite{yule2014statistical} seems to be almost independent of text length.
Given a text of length $N$ with $V$ unique words, Yule's K is defined as
\begin{equation}
    K = 10^4 \cdot \left[-\frac{1}{N} + \sum_{i = 1}^{V} V(i,N) \left(\frac{i}{N}\right) ^2 \right]
\end{equation}
where $V(i,N)$ denotes the number of words appearing $i$ times in the text. In particular, the larger is $K$, the less rich the vocabulary is.

For what concerns the repetitiveness of text, we use an approach akin to previous works \cite{parada2024song,desiderio2023recurring}, i.e. we compress the raw texts using {\it gzip} and compare their dimensions with the original ones. We define the {\it gzip} complexity $g$ as 
\begin{equation}
    g = \frac{s_{raw} - s_{compressed}}{s_{raw}}
\end{equation}
where $s_{raw}$ is the size of the raw text and $s_{compressed}$ is the size of the compressed text.
Note that if the text is highly repetitive $s_{compressed} \ll s_{raw}$ and therefore $g \approx 1$. On the other hand, low values characterize texts with low repetitiveness. In particular, for very short texts the compressor may increase the size, therefore $g$ can also assume negative values. 

\subsection*{Classification of users}
We adopt a non-parametric method developed by~\cite{citation_impact} to partition heavy-tailed distributions, and recently employed in different domains~\cite{abramo2017investigation,cinelli2021ambiguity,DiMarco2024} to divide users into classes according to the number of comments they left. 
In detail, we consider four classes of activity, namely $\{low, mid, high, very \, high\}$, and adopt the following procedure: firstly, we compute the mean number of comments $\Bar{x}$ and assign to class {\it low} all users that have left less than $\Bar{x}$ comments; then, we delete these users from the distribution and recursively repeat the procedure until each user is assigned to one of the four classes $\{low,mid,high,very \; high\}$.

\subsection*{Regression model}
Since we are interested in obtaining a unique model capable of detecting the overall relationship between time, social media and the complexity of comments, we consider

\begin{equation}
    y_i \sim [1 \; w_i \; K_i \; g_i] \cdot {\bf B} \cdot 
    \begin{bmatrix}
        1 \\ tw_i \\ vt_i \\ yt_i \\ un_i
    \end{bmatrix} 
\end{equation}

where $\cdot$ is the standard matrix product and 

\begin{itemize}
    \item $w_i$ is the number of types of comment $i$;
    \item $K_i$ is the $K-$complexity of comment $i$;
    \item $g_i$ is the $g$-complexity of comment $i$;
    \item $y_i$ is the year in which comment $i$ has been created;
    \item $tw_i,vt_i,yt_i$ and $un_i$ are dummy variables that are equal to one if and only comment $i$ has been written in Twitter, Voat, YouTube or Usenet, respectively.
\end{itemize}

Finally, ${\bf B}$ is the matrix of the coefficients, defined as follows:
$$
B = 
\begin{bmatrix}
    \beta_0 & \beta_{0,tw} & \beta_{0,vt} & \beta_{0,yt} & \beta_{0,un} \\ 
    \beta_1 & \beta_{1,tw} & \beta_{1,vt} & \beta_{1,yt} & \beta_{1,un} \\ 
    \beta_2 & \beta_{2,tw} & \beta_{2,vt} & \beta_{2,yt} & \beta_{2,un} \\ 
    \beta_3 & \beta_{3,tw} & \beta_{3,vt} & \beta_{3,yt} & \beta_{3,un} \\ 
\end{bmatrix}
$$

We use Facebook as the baseline, i.e. the estimates of coefficients $\beta_i$ refer to Facebook comments. Therefore, the generic coefficient $\beta_k + \beta_{k,j}$ describes the change in $y$ due to a unit increase of regressor $k$ if all the other variables are kept constant and comment $i$ comes from a social $j$ different from Facebook. 
To obtain more interpretable estimates, we first normalize all the regressors. Therefore, the parameters quantify changes in the dependent variable in standard deviations. 
Moreover, since we detect heteroskedasticity, we correct the errors and tests using the {\it sandwich} package \cite{sandwich,sandwich_lm} in R.

\section*{Acknowledgements}
The work is supported by IRIS Infodemic Coalition (UK government, grant no. SCH-00001-3391), 
SERICS (PE00000014) under the NRRP MUR program funded by the European Union - NextGenerationEU, project CRESP from the Italian Ministry of Health under the program CCM 2022, PON project “Ricerca e Innovazione” 2014-2020, and PRIN Project MUSMA for Italian Ministry of University and Research (MUR) through the PRIN 2022. 
This work was supported by the PRIN 2022 “MUSMA” - CUP G53D23002930006 - Funded by EU - Next-Generation EU – M4 C2 I1.1.

\section*{Supplementary Information}
\subsection*{Type-Token ratio of users' comments}
\begin{figure*}[!ht]
    \centering
    \includegraphics[width = .8\linewidth]{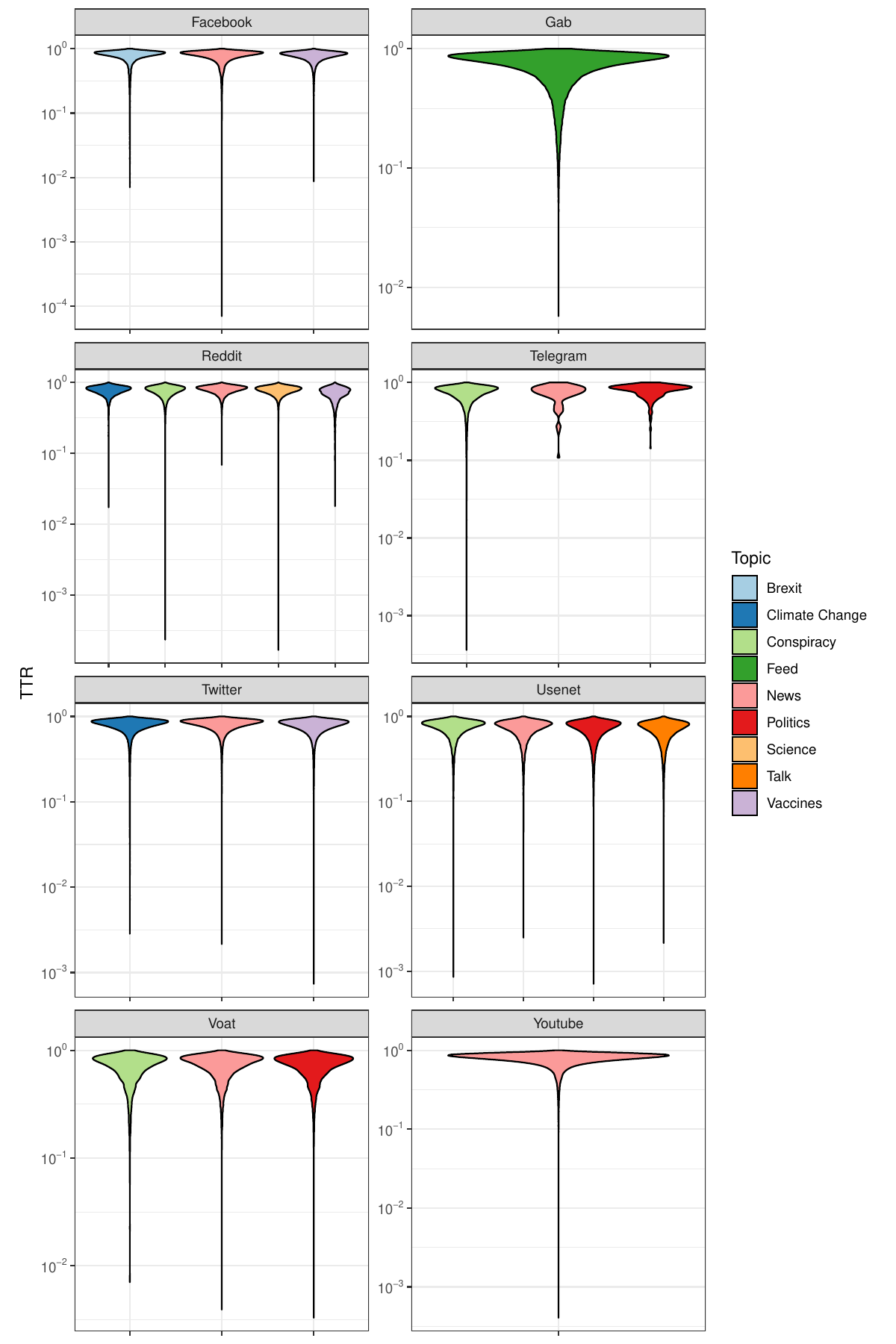}
    \caption{Distribution of TTR values of comments of users. To consider consistent cases, we only compute TTR for users having at least 50 tokens.}
    \label{fig:TTR_dist}
\end{figure*}

We used the Type-Token ratio (TTR), a simple measure defined as the ratio of unique words over the total words of a text, to evaluate the lexical complexity of users' comments. Figure \ref{fig:TTR_dist} shows the distribution of TTR values obtained from the documents created with text of each user.

The distribution are mostly peaked at one, thus indicating a high similarity between the number of types and tokens.

\subsection*{Zipf's law on comments}
Zipf's law was initially designed and studied in \cite{zipf2013psycho,zipf2016human} to model the frequency at which words are observed in texts. In particular, it simply states that the $r-$th most frequent word has frequency $f(r)$ that scales according to

\begin{equation}\label{eq:zipf}
  f(r) \approx r^{-\alpha}
\end{equation}

In particular, real texts often show $\alpha \approx 1$.

Although many alternatives have been proposed (see \cite{mandelbrot1953informational, mandelbrot1961theory} for example) and some works suggest that the right form decay of $f(r)$ may depend on the context \cite{Piantadosi2014-vo}, here we rely on the simplest form of \eqref{eq:zipf}.

\begin{figure}[!ht]
    \centering
    \includegraphics[width = .9\linewidth]{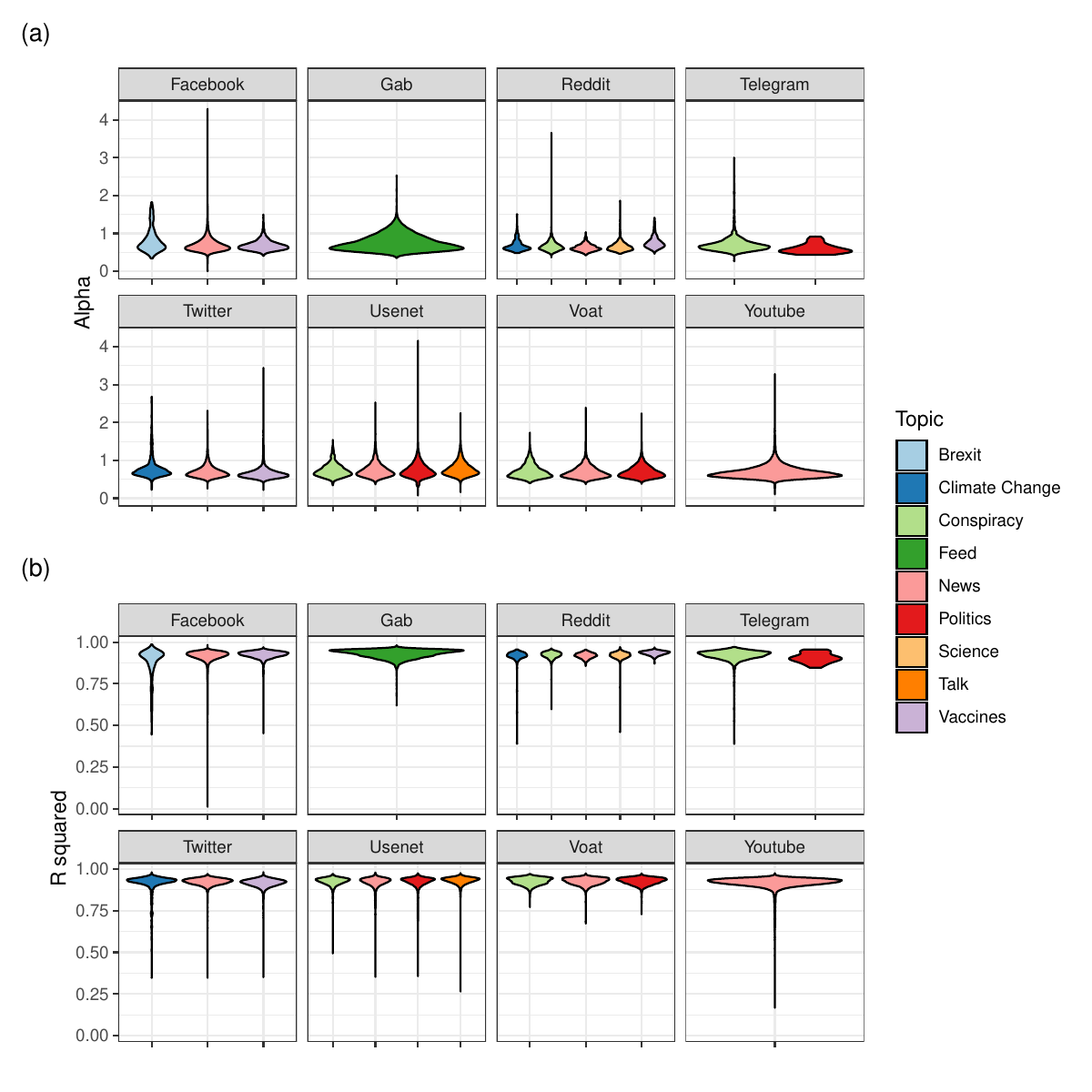}
    \caption{Distribution of the $(a)$ exponent of the Zipf's Law and $(b)$ $R^2$ of the regressions. Telegram News was excluded since no user had more than 1000 tokens.}
    \label{fig:zipf}
\end{figure}

In particular, we aim to understand if it is reasonable to suppose that the comments generated by users follow \eqref{eq:zipf}. 
As explained in the main text, for each user we paste together her/his generated texts, obtaining a document for each of them.
We then apply the standard tokenization and stemming procedure and keep only users with at least 1000 tokens. In this way, we select a subset of active users for which the fitting may be reliable. 
We then apply a linear regression to the logged values of $r$ and $f(r)$ and store the results. 
Figure \ref{fig:zipf} $(a)$ shows the distribution of $\alpha$ for all the significant fit ($p < 0.001$). Note that Telegram News was not included since no users have a sufficient number of tokens. 

Interestingly, we observe very similar distributions centered in $\approx 0.7$. Moreover, the majority of regressions explain a very large fraction of the variance, as depicted in Figure \ref{fig:zipf} $(b)$.
The results thus suggest that equation  \eqref{eq:zipf} may be a reliable approximation for the frequency of terms used by online users, even if the exponents show lower values than the ones commonly observed. 

\section*{Null model of complexity measures}
To detect if each user possesses her/his distinct vocabulary, we select and randomize the comments of users (with more than $20$ comments) before the aggregation into documents in each distinct dataset. Doing that, each user is associated with a set of random comments, having the same cardinality as the original document.

To ensure consistent statistics and manageability, we select a sample of $15K$ users for bigger datasets.

\begin{figure*}[!ht]
    \centering
    \includegraphics[width = .9\linewidth]{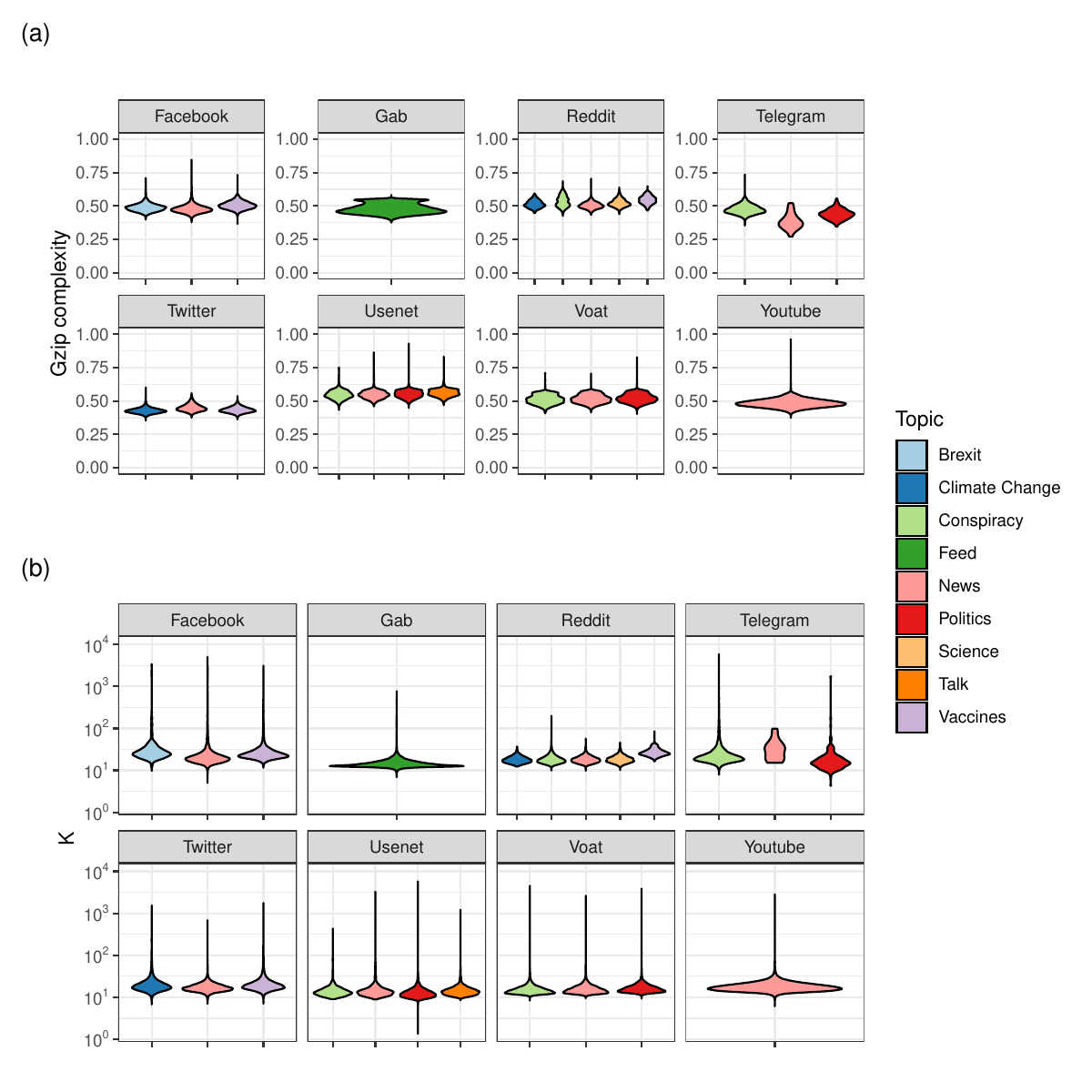}
    \caption{Distribution of {\it gzip} complexity and Yule's $K$ over documents obtained in the case in which the comments of users are randomized.}
    \label{fig:null_model}
\end{figure*}

We then compute {\it gzip} complexity and Yule's $K$ for each document. 
The resulting distributions are shown in Figure \ref{fig:null_model}.

We observe narrower distributions than the real case, suggesting even more uniform behaviours. To check if statistically significant differences arise between the real and the null distributions, we apply $2-$sided Mann-Whitney tests, whose results are shown in Table \ref{tab:mann_whitney}. Interestingly, we observe that, in almost all cases, the distributions are significantly different thus suggesting that each user has her/his distinct vocabulary that is not possible to reply with a random assignment of texts.

\subsection*{Logistic regression}
In this section, we define a logistic regression model employing the same data used in the main paper (Table \ref{tab:breakdown_regression} shows its data breakdown).

Let's consider the dataset comprising the comments from social $j$ and let $d$ be its distribution of years. 
We say that a comment $i$ is {\it recent} if $year(i) > median(d)$.
Then, we employ a logistic regression model to predict if a comment is recent based on their measures of complexity.
More in detail, the model has the following form:

\begin{equation}
    p = \frac{1}{1 + e^{-z}},
\end{equation}

or equivalently

\begin{equation}
    z = \ln \frac{p}{1-p},
\end{equation}

where $z = \beta_0 + \beta_1 w + \beta_2 K + \beta_3 g$ and $w,K,g$ are the number of types, the $K$-complexity and $g-$complexity respectively.

Thus, a positive estimate of parameter $\beta_j, j = 1,2,3$ indicates that an increase in complexity $j$ increases the logit of being a {\it recent} comment. The opposite result holds if $\beta_j$ is negative. 

We construct a model for each social separately. The results of the analysis are reported in Table \ref{tab:logistic_regression}, where we show only the values of the parameter for each social and their significativity.

Although not all the coefficients are significant, we can see a general coherence with the result obtained by the model presented in the main paper, with the only differences arising for Twitter. 

Moreover, by applying the same methodology to the whole dataset without distinguishing for the social of provenience, we obtain the following estimates (all significant at the $0.001$ level): $\beta_0 = 0.744, \beta_1 = -0.028, \beta_2 = 2.489 \cdot 10^{-4}$ and $\beta_3 = -1.044$, coherent with our results.

Overall, the analysis confirms that low values of types and $g$ are linked to recent comments, while low values of $K$ are linked to less recent texts.

\clearpage
\subsection*{Supplementary Figures}

\clearpage

\clearpage

\clearpage

\clearpage
 
\clearpage

\subsection*{Supplementary tables}

\begin{table}[!ht]
    \centering
        \caption{Data breakdown of the dataset used in the regression analysis}
\begin{tabular}{llrrrr}
  \toprule
Social & Topic & Min year & Max year & Comments & Users \\ 
  \midrule
Facebook & News & 2010 & 2016 & 42000 & 39096 \\ 
  Facebook & Vaccines & 2010 & 2017 & 48000 & 27742 \\ 
  Twitter & Vaccines & 2015  & 2022  & 36000 & 33042 \\ 
  Usenet & News & 1994  & 2005  & 72000 & 21103 \\ 
  Usenet & Politics & 1994  & 2005  & 72000 & 26739 \\ 
  Usenet & Talk & 1992  & 2005  & 84000 & 27416 \\ 
  Voat & News & 2015  & 2020  & 36000 & 9727 \\ 
  Voat & Politics & 2015  & 2020  & 36000 & 8991 \\ 
  Youtube & News & 2008  & 2022  & 90000 & 77682 \\ 
   \bottomrule
\end{tabular}

    \label{tab:breakdown_regression}
\end{table}
\clearpage 

\begin{table}[!ht]
\centering
\caption{Results of Mann-Whitney tests comparing $g$ and $K$ distributions from real and null model case. }
\begin{tabular}{lrr}
  \toprule
Dataset & $p_g$ & $p_k$ \\ 
  \midrule
Facebook Brexit &  <0.001 &  <0.001 \\ 
  Facebook Vaccines &  <0.001 &  <0.001 \\ 
  Gab Feed &  <0.001 &  <0.001 \\ 
  Reddit Climate Change &  <0.001 &  <0.001 \\ 
  Reddit Conspiracy &  <0.001 &  <0.001 \\ 
  Reddit News &  <0.001 &  <0.001 \\ 
  Reddit Science &  <0.001 &  <0.001 \\ 
  Reddit Vaccines &  <0.001 &  <0.001 \\ 
  Telegram Conspiracy &  <0.001 &  <0.001 \\ 
  Telegram News & 0.650 & 0.001 \\ 
  Telegram Politics & 0.717 &  <0.001 \\ 
  Twitter Climate Change &  <0.001 &  <0.001 \\ 
  Twitter News &  <0.001 &  <0.001 \\ 
  Twitter Vaccines &  <0.001 &  <0.001 \\ 
  Usenet Conspiracy &  <0.001 &  <0.001 \\ 
  Usenet News &  <0.001 &  <0.001 \\ 
  Usenet Politics &  <0.001 &  <0.001 \\ 
  Usenet Talk &  <0.001 &  <0.001 \\ 
  Voat Conspiracy &  <0.001 &  <0.001 \\ 
  Voat News &  <0.001 &  <0.001 \\ 
  Voat Politics &  <0.001 &  <0.001 \\ 
  Youtube News &  <0.001 &  <0.001 \\ 
  Facebook News &  <0.001 &  <0.001 \\ 
   \bottomrule
\end{tabular}
\label{tab:mann_whitney}
\end{table}
\clearpage

\begin{table}[!ht]
    \centering
        \caption{Result of logistic regression for each dataset. $*,**,***$ indicate that the coefficient is significant at $0.05,0.01,0.001$ level respectively.}
    \begin{tabular}{lcccc}
    \toprule
       Social & $\beta_0$ & $\beta_1$ & $\beta_2$ & $\beta_3$ \\
       \midrule
       Usenet & $-0.103^{***}$ & $-0.013^*$ & $0.015^{***}$ & $-0.369^{***}$ \\
       
       Facebook & $-0.132^{***}$ & $-0.096^{***}$ & $0.084^{***}$ & $-0.298^{***}$ \\
       YouTube & $-0.133^{***}$ & $0.069^{***}$ & $0.004$ & $-0.026^{**}$ \\
              Voat & $0.001$ & $-0.006$ & $0.007$ & $-0.164^{***}$ \\
       Twitter & $-0.275^{***}$ & $0.029^{***}$ & ${-2.047 \cdot 10^{-4}}^{***}$ & $0.036$ \\
       \bottomrule
    \end{tabular}

    \label{tab:logistic_regression}
\end{table}
\clearpage

\clearpage
\bibliography{bibliography}
\bibliographystyle{plain}
\end{document}